\documentclass[reprint,superscriptaddress,aps,prd,floatfix]{revtex4-2}
\usepackage[a4paper,left=1.5cm,right=1.5cm,top=3cm,bottom=3cm]{geometry}
\usepackage[toc]{appendix} 

\usepackage{graphicx}
\usepackage{dcolumn}
\usepackage{bm}
\usepackage{amssymb,amsmath,amsfonts,physics}
\usepackage[dvipsnames]{xcolor}
\usepackage{graphicx}
\usepackage{longtable}
\usepackage{verbatim}
\usepackage{color}
\usepackage{mdframed}
\usepackage{soul}
\usepackage{amsfonts,amssymb,mathrsfs,amsmath,esint}
\usepackage{slashed, cancel}
\usepackage{framed}
\usepackage{mdframed}
\usepackage{simplewick} 
\allowdisplaybreaks
\usepackage{latexsym}
\usepackage{graphicx}
\graphicspath{{./Figures/}}
\usepackage[dvipsnames]{xcolor}
\usepackage{booktabs}
\usepackage{datetime}
\newdateformat{mydate}{\THEDAY{ }\monthname[\THEMONTH]{ }\THEYEAR}

\usepackage{tikz}
\usepackage{color}
\usepackage{framed}
\usepackage{hyperref}
\usepackage[capitalise]{cleveref}
\hypersetup{colorlinks, citecolor=bluscuro, linkcolor=black, urlcolor=bluscuro}
\definecolor{rossos}{cmyk}{0,1,1,0.55}
\definecolor{bluscuro}{rgb}{0.15, 0.2, .85}
\definecolor{bluchiaro}{cmyk}{1,.3,0.,0.1}

\graphicspath{{./images/}}

\newcommand{\be}{\begin{equation}}
	\newcommand{\ee}{\end{equation}}
\newcommand{\bea}{\begin{eqnarray}}
	\newcommand{\eea}{\end{eqnarray}}
\newcommand{\beq}{\begin{equation}}
	\newcommand{\eeq}{\end{equation}}

\def\beqa{\begin{eqnarray}}

	\def\eeqa{\end{eqnarray}}

\def\lsim{\mathrel{\rlap{\lower4pt\hbox{\hskip0.5pt$\sim$}}
		\raise1pt\hbox{$<$}}}         
\def\gsim{\mathrel{\rlap{\lower4pt\hbox{\hskip0.5pt$\sim$}}
		\raise1pt\hbox{$>$}}}         

\usepackage[normalem]{ulem}
\usepackage{soul}

\begin{document}
	
	\title{Evidence for non-cold dark matter from DESI DR2 measurements}
	
	\author{Utkarsh Kumar}
	\email{utkarshkumar.physics@gmail.com}
	\affiliation{Department of Physics, University of Ottawa, Ottawa, ON K1N6N5, Canada;}
	\author{Abhijith Ajith}
	\email{abhijith.ajith.421997@gmail.com}
	\affiliation{Indian Institute of Science Education and Research Bhopal,
Bhopal 462066, India}
	\author{Amresh Verma}
	\email{amreshverma702@gmail.com}
	\affiliation{Physics Department, Ariel University, Ariel 40700, Israel}
	\date{\today}

\begin{abstract}
We investigate potential deviations from cold dark matter (CDM) using the latest Baryon Acoustic Oscillations (BAO) measurements from the Dark Energy Spectroscopic Instrument (DESI). Analyzing DESI data alone constrains the dark matter equation of state parameter $w_{\mathrm{dm}} = -0.042^{+0.047}_{-0.024}$, revealing a mild preference for non-cold dark matter. This preference strengthens significantly in combined analyses, but reveals a striking tension in the inferred $w_{\mathrm{dm}}$ values from observations of different nature. The DESI+DESY5 combination yields $w_{\mathrm{dm}} = -0.084 \pm 0.035$, excluding CDM ($w_{\mathrm{dm}}=0$) at 2.4$\sigma$ significance. In contrast, Planck+DESI gives $w_{\mathrm{dm}} = 0.00077\pm0.00038$, differing from concordance model at 2$\sigma$ significance. The non-vanishing $w_{\mathrm{dm}}$ preference is particularly driven by low-redshift BAO measurements ($z<1.1$), while higher redshift data remain consistent with $\Lambda$CDM. The evidence for non-cold dark matter is more pronounced in DESI compared to the previous BAO surveys. All dataset combinations show significant improvement over the $\Lambda$CDM paradigm, providing compelling evidence for non-cold dark matter scenario.
\end{abstract}

	\maketitle
	
	\textbf{Introduction}-- 
    The standard model of cosmology explains the numerous phenomena of the observed Universe while providing an excellent fit to the various astrophysical and cosmological observations. The measurements from the cosmic microwave background (CMB) \cite{Planck:2018vyg,Planck:2019nip,Planck:2018lbu, SPT-3G:2021eoc, ACT:2020frw, WMAP:2012nax, Planck:2018nkj,Lemos:2023rdh,Tristram:2007zz}, baryon acoustic oscillations (BAO) \cite{BOSS:2016wmc,eBOSS:2020tmo,eBOSS:2020gbb,eBOSS:2020yzd,eBOSS:2020hur,eBOSS:2020lta,DESI:2025fxa,DESI:2024mwx}, Type Ia supernovae \cite{Pan-STARRS1:2017jku,Scolnic:2021amr,Rubin:2023ovl,Bailey:2022pax,Blondin:2024fpr,DES:2024jxu,NearbySupernovaFactory:2015pcf} and weak lensing \cite{DES:2016jjg,DES:2018ekb,LineaScienceServer:2021mgv,Hildebrandt:2016iqg,Hildebrandt:2018yau,HSC:2018mrq,DES:2022ccp,DES:2021wwk}, have significantly improved our understanding of the Universe. The standard model of cosmology suggests that most of the universe comprises two unknown components, called dark matter (DM) and dark energy (DE), which are responsible for structure formation and the current accelerated expansion of the Universe, respectively. Several cosmological observations such as Supernovae type Ia measurements and the more recent BAO measurements from the second data release (DR2) of the Dark Energy Spectroscopic Instrument (DESI) have shown strong evidence towards the existence of the Universe's accelerated expansion. On the other hand, in the absence of any direct observations of dark matter, we are only left to presume about its property based on its weak interactions with standard model particles and gravitational effects. In the standard cosmological scenario, due to its weak interaction with other known particles, the dark matter component is parameterised as a pressure-less fluid with vanishing equation of state (EoS), referred to as cold dark matter(CDM).

CDM explains the formation of structures consistently at scales much higher than 1 Mpc. Nevertheless, it encounters issues while looking at observations on smaller scales \cite{Copi:2013cya, Copi:2006tu,Bullock:2017xww}
     like the core-cusp \cite{deVega:2013jfy}, missing satellite \cite{Klypin:1999uc,Moore:1999nt} and the too-big-to-fail problems \cite{2011MNRAS.415L..40B}.
     Such discrepancies have prompted the consideration of alternative dark matter models. Warm dark matter(WDM) composed of particles with non-negligible thermal velocities offers a viable resolution to these problems. Unlike CDM, WDM suppresses structure formation below its free streaming scale. Hence, WDM models erase the substructure at smaller scales, delay the structure formation, and resolve the small-scale problems that plague the standard cosmology. Sterile neutrinos are said to be one of the viable candidates of warm dark matter \cite{Dodelson:1993je, Lovell:2015psz}.

The strong preference for evolving dark energy from DESI \cite{DESI:2024mwx,Cortes:2024lgw}, has prompted the surge in the exploration of a plethora of models of DE \cite{PhysRevD.37.3406,Chiba:1997ej,PhysRevD.62.023511,PhysRevD.63.103510,Sotiriou:2008rp,Manoharan:2022qll,Rinaldi:2014yta,Hu:2007nk,Casalino:2018tcd,Mukhopadhyay:2019cai,Ruchika:2020avj,Odintsov:2020zct,Ong:2022wrs,Luciano:2023wtx,Tyagi:2025zov,Akrami:2025zlb,Santos:2025wiv,Colgain:2025nzf,Artymowski:2020zwy,Ben-Dayan:2023rgt,Ben-Dayan:2023htq}, and DE-DM interactions \cite{Amendola:1999er,An:2017crg,Yang:2017yme,DiValentino:2019ffd,Patil:2023rqy,vanderWesthuizen:2023hcl,Ashmita:2024ueh, Gomes:2023dat}. In this letter, we use the DESI measurements to probe the nature of dark matter. In the absence of any underlying theory of dark matter, we investigate a phenomenological model with a dark matter component of an arbitrary EoS in the light of recent data release of DESI measurements. We demonstrate that the DESI data give a robust preference for the non-vanishing pressure of dark matter candidate.

    \textbf{Cosmology with non-cold dark matter}-- We consider the Universe to be filled with a dark matter component of energy density $\rho_{\rm dm}$ and EoS parameter $w_{\rm dm}$ \cite{Kumar:2012gr, Muller:2004yb,Armendariz-Picon:2013jej, Kopp:2018zxp}. The evolution of the energy density of such a dark matter component, determined by solving the continuity equation, scales as
    \begin{equation}
        \rho_{\rm dm} = \rho_{\text{dm},0}\, a^{-3\left(1 + w_{\rm dm}\right)}
    \end{equation}
    with $\rho_{\rm dm,0}$ being the current energy density and $a$ being the scale factor. In the presence of $\rho_{\rm dm}$, the Friedmann equations take the following form:
    \begin{align}
        \mathcal{H}^{2} & = \frac{8 \pi G a^2}{3} \, \left( \rho_{\rm r} + \rho_{\rm b} +   \rho_{\rm dm} + \rho_{\rm \Lambda}\right) \\ 
        \dot{\mathcal{H}}-\mathcal{H}^2  & = - 4 \pi G a^2 \, \left( \frac{4}{3}\,\rho_{\rm r} + \rho_b + \left(1 + w_{\rm dm}\right)\,\rho_{\rm dm}\right)
    \end{align}
    Here $\rho_{\rm r,b}$ are the energy densities of radiation and baryonic components of the universe which scale as $\rho_{\rm r,0}\, a^{-4}$ and $ \rho_{\rm b,0}\, a^{-3} $ respectively. Note that the derivatives considered here are with respect to conformal time and $\mathcal{H}$ is the corresponding Hubble parameter. Due to the non-vanishing pressure of dark matter, the perturbation equations also get modified, and are expressed as follows in the synchronous gauge without the inclusion of shear perturbations \cite{Ma:1995ey,Xu:2013mqe}:
    \begin{align}
        \begin{split}
            \Dot{\delta}_{\rm dm} &= -\left(1 + w_{\rm dm}\right) \, \left(\theta_{\rm dm} + \Dot{h} / 2\right) - 3\,\mathcal{H} \left(c_{\rm s}^{2} - c_{\rm a}^{2}\right)  \\
            & \left[ \delta_{\rm dm} + 3\,\mathcal{H} \left(1 + w_{\rm dm}\right)\, \frac{\theta_{\rm dm}}{k^{2}}\right]
        \end{split} \\
\begin{split}
            \Dot{\theta}_{\rm dm} &= -\mathcal{H}\,\left(1 - 3 c_{\rm s}^{2}\right)\,\theta_{\rm dm} + \frac{k^{2}  c_{\rm s}^2}{1 + w_{\rm dm}}\, \delta_{\rm dm}
        \end{split}
    \end{align}
    Here $\delta_{\rm dm}$ and $\theta_{\rm dm}$ represent the density contrast and velocity divergence of dark matter, respectively. We model dark matter to be a perfect fluid with a constant equation of state, in which the definition of adiabatic speed of sound becomes $c_{\rm a}^{2} = w_{\rm dm}$. Further, we assume the effective speed of sound $c_{\rm s}^2$ to be 0.
    
\begin{figure*}[!ht]
    \centering
    \includegraphics[scale=0.53]{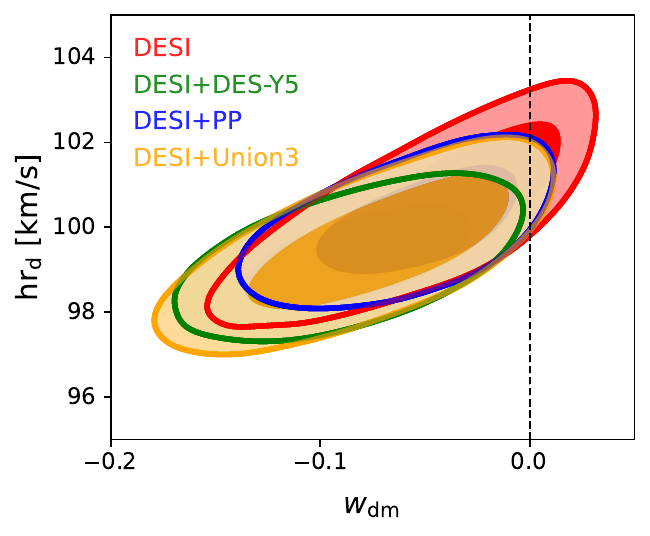}
    \includegraphics[scale=0.53]{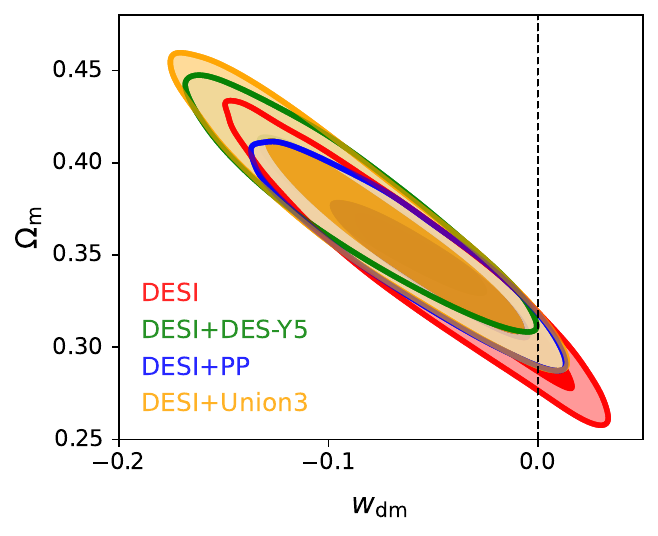}
    \includegraphics[scale=0.50]{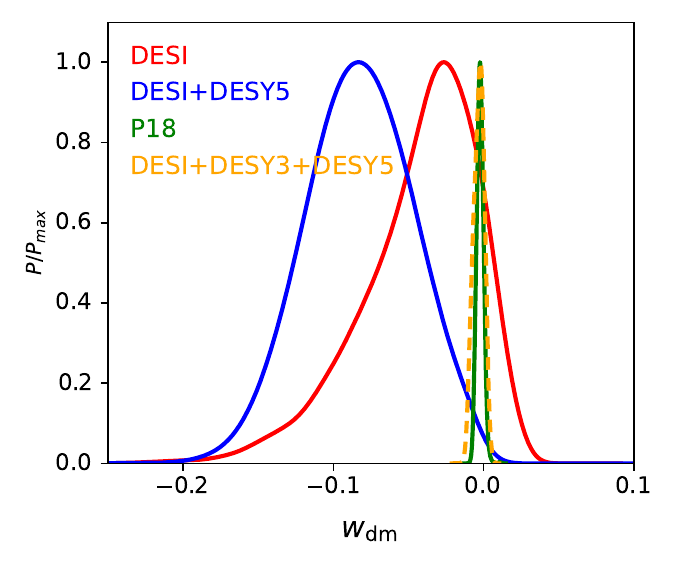}
    \caption{Two-dimensional confidence contours (68\% and 95\% CL) for the cosmological parameter pairs $h\,r_{\mathrm{d}}$--$w_{\mathrm{dm}}$ (left panel) and $\Omega_{\mathrm{m}}$--$w_{\mathrm{dm}}$ (middle panel), derived from \textsc{DESI} and supernovae (SN) datasets. The vertical black dashed line indicates the cold dark matter equation of state ($w_{\mathrm{dm}} = 0$). Marginalized constraints on $w_{\mathrm{dm}}$ are shown for CMB (P18), DESI BAO, SNIa (DES-Y5), and LSS (DES-Y3). Color coding for each dataset corresponds to the legend.}
    \label{fig:wdm_posteriors}
\end{figure*}

\textbf{Methodology and data}-- We implement the theoretical model in a modified version of the Boltzmann solver code \texttt{CAMB} \cite{Lewis:1999bs,2012JCAP...04..027H}. We perform Markov-Chain-Monte Carlo (MCMC) simulations using the publicly available tool \texttt{COBAYA} \cite{Torrado:2020dgo,2019ascl.soft10019T}. Our model has 7 free cosmological parameters. We constrain the cosmological parameters with uniform priors: the baryonic matter density $\Omega_b h^{2} \in [0.005, 0.1]$, the dark matter density $\Omega_{\rm dm}h^{2} \in [0.001,0.99]$, the logarithmic amplitude of primordial curvature spectrum $ {\rm{ln}}(10^{10} A_s) \in [1.6,3.9]$ evaluated at a suitable pivot scale, $k = 0.05 \  \text{Mpc}^{-1}$ along with its tilt $ n_s \in [0.8,1.2] $, the reionization optical depth $\tau_{\rm reio} \in [0.01,0.8] $, the present value of the Hubble parameter $ H_0 \in [20,100] $ and EoS for dark matter $w_{\rm dm} \in [-0.5,0.5]$. We use the standard three-neutrino description with a massive neutrino of mass $m_{\nu}$ = 0.06 eV, while the other two being  massless. The chains converge by ensuring the Gelman-Rubin criterion $|R-1| \leq 0.01$ or the effective sample size becoming greater than $10^5$. We utilize  \texttt{GetDist} \cite{Lewis:2019xzd} and \texttt{BOBYQA} \cite{2018arXiv180400154C,2018arXiv181211343C} to analyze chains and find the maximum likelihood posterior $\chi^2_{\rm MAP}$, respectively. We calculate the differences in $\chi^2_{\rm MAP}$ and deviance information criterion (DIC) to get the preference of our model over $\Lambda$CDM. The DIC is calculated as 
\begin{equation}
    \rm DIC = 2 \overline{\chi^{2}\left(\theta\right)} - \chi^{2}(\hat{\theta})
\end{equation}
with $\overline{\chi^{2}\left(\theta\right)}$ and $\chi^{2}(\hat{\theta})$ being the average of the effective $\chi^2$ over the posterior distribution and the best-fit $\chi^2$ respectively. In addition, we also estimate the concordance/discordance between two datasets, $D_1$ and $D_2$ in a given model using the DIC-based statistic $\mathcal{I}\left(D_1,D_2\right)$ \cite{Vagnozzi:2020rcz}.
To be precise, a positive value of $\log_{10}\mathcal{I}$ indicates the consistency between the two datasets and conversely for a negative value. We further qualify the level of concordance or discordance using Jeffrey's scale, such that the value of $\log_{10}\mathcal{I}$ in excess of $\pm1/2$ is considered to be `substantial',  excess of $\pm1$ is `strong', and a value in excess of $\pm2$ shall be `decisive' \cite{Joudaki:2016mvz}.    
    We use the following  datasets in our analysis:
    \begin{itemize}
        \item \textbf{DESI:-} The 13 DESI-BAO DR2 (\cite{DESI:2025zgx,DESI:2025zpo}) measurements across the redshifts range $0.1 < z < 4.2$ obtained from observations of about 14 million galaxies and quasars which include bright galaxy sample (BGS), luminous red galaxies (LRG), emission line galaxies (ELG), quasars (QSO), and Lyman-$\alpha$ tracers. These measurements are given in terms of the volume averaged distance $D_{\rm V}(z) / r_{\rm d}$, angular diameter distance $D_{\rm M}(z) / r_{\rm d}$ and comoving Hubble distance $D_{\rm H}(z) / r_{\rm d}$, where $r_{\rm d}$ is the sound horizon at the drag era.
        \item \textbf{CMB:-} CMB temperature$-$temperature, temperature -polarization, and polarization-polarization spectra \cite{Planck:2018vyg}, along with the CMB lensing likelihood derived from the 4-point correlation function of the CMB \cite{Planck:2018lbu}. We refer to these as P18 and PL18, respectively.

        \item \textbf{SNIa:-} We use the PantheonPlus(PP) dataset \cite{Scolnic:2021amr}, which contains 1701 light curves for 1550 spectroscopically confirmed Type Ia supernovae (SNeIa) covering the redshift range $0.001 < z < 2.26$. Additionally, we incorporate the Union3 compilation comprising 2087 SNe \cite{Rubin:2023ovl} and the DES-Y5(DESY5) sample containing 1635 DES SNe spanning $0.10<z<1.13$ \cite{DES:2024jxu}.

        \item \textbf{DESY3 3×2pt:-} We use galaxy-galaxy, galaxy-shear, and shear-shear two-point correlation function measurements  obtained from DES Year 3 analysis. These observations were measured from over 10 million lens galaxies in the \texttt{MagLim} sample, that covers approximately 4000 $\text{deg}^2$ of sky \cite{DES:2022ccp,DES:2021wwk}. We use the \textsc{CosmoLike} \cite{Krause:2016jvl} pipeline in our analysis of DESY3.

    \end{itemize}

\begin{table*}[!ht]
\centering
\label{tab:constraints}
\setlength{\tabcolsep}{4pt}
\renewcommand{\arraystretch}{1.0}
\begin{tabular}{|l|c|c|c|c|c|}
\hline
\hline
Dataset & $\bm{H_0}$ & $\bm{w_{\rm dm}} \, (\# \, \sigma)$ & $\bm{\Omega_{m}}$ & $\bm{h\,r_{\mathrm{d}}}$ & $\bm{S_{8}}$ \\
\hline
\hline
DESI & $80^{+20}_{-7}$ & $-0.042^{+0.047}_{-0.024}$ (1.09)& $0.334^{+0.023}_{-0.040}$ & $100.5^{+1.2}_{-1.0}$ & -- \\
DESI+DESY5 & $75^{+10}_{-8}$ & $-0.084 \pm 0.035$ (2.38)& $0.375^{+0.026}_{-0.032}$ & $99.27 \pm 0.81$ & -- \\
DESI+PP & $78^{+10}_{-9}$ & $-0.055^{+0.038}_{-0.025}$ (1.71)& $0.346^{+0.021}_{-0.030}$ & $100.09 \pm 0.84$ & -- \\
DESI+Union3 & $77^{+10}_{-10}$ & $-0.074^{+0.049}_{-0.034}$ (1.79)& $0.365^{+0.027}_{-0.043}$ & $99.6^{+1.1}_{-0.98}$ & -- \\
\hline
PL18 & $63.1 \pm 3.9$ & $-0.0022 \pm 0.0020$ (1.09)& $0.383^{+0.051}_{-0.073}$ & $91.8 \pm 6.9$ & $0.867 \pm 0.035$ \\
PL18+DESI & $69.03 \pm 0.42$ & $0.00077 \pm 0.00038$ (2.01)& $0.2947 \pm 0.0047$ & $101.63 \pm 0.60$ & $0.8202 \pm 0.0099$ \\
PL18+DESI+DESY5 & $68.63 \pm 0.40$ & $0.00054 \pm 0.00037$ (1.45)& $0.2993 \pm 0.0046$ & $101.06 \pm 0.58$ & $0.8213 \pm 0.0099$ \\
PL18+DESI+PP & $68.81 \pm 0.40$ & $0.00064 \pm 0.00037$ (1.76)& $0.2972 \pm 0.0046$ & $101.32 \pm 0.58$ & $0.8209 \pm 0.0098$ \\
PL18+DESI+Union3 & $68.85 \pm 0.42$ & $0.00067 \pm 0.00037$ (1.80)& $0.2967 \pm 0.0047$ & $101.38 \pm 0.59$ & $0.8210 \pm 0.0099$ \\
\hline
DESY3 (3$\times$2pt) & $ 82^{+20}_{-6}$ & $ -0.0087^{+0.0067}_{-0.0058}$ (1.46)& $ 0.377^{+0.044}_{-0.066}$ & $100.7^{+4.9}_{-7.1}$ & $0.766 \pm 0.027$\\
DESI+DESY3 (3$\times$2pt) & $>80$ & $-0.0005^{+0.0035}_{-0.0026}$ (0.16)& $0.3007^{+0.0094}_{-0.011}$ & $101.32 \pm 0.76$ & $0.801 \pm 0.017$ \\
DESI+DESY3 +DESY5 & $>81.4$ & $-0.0025^{+0.0040}_{-0.0029}$ (0.73)& $0.3148 \pm 0.0094$ & $100.35 \pm 0.68$ & $0.794^{+0.017}_{-0.016}$ \\
\hline
\hline
\end{tabular} 
\caption{The mean $\pm 1 \sigma$ constraints on the cosmological parameters inferred from the various datasets and their combinations considered in this letter for the $\Lambda$wDM scenario. In the bracket next to the $w_{dm}$ we report the sigma deviation of $w_{dm}$ from its expected value, zero. To calculate this, we compute the standard deviation $\sigma_{w_{\text{dm}}}$  from the covariance matrix and tension is defined as the deviation of the mean ($\mu_{w_{\text{dm}}}$) from zero in units of the standard deviation: $\frac{|\mu_{w_{\text{dm}}}|}{\sigma_{w_{\text{dm}}}}$. We note  $\sim 2.4 \, \sigma$ preference for $w_{\rm dm}<0$ in DESI+DESY5 and discrepancy between DESI and Planck measurements.}
\label{tab:1sigma}
\end{table*}

The $68\%$ constraints on the cosmological parameters for various datasets considered in this letter are shown in \cref{tab:1sigma}. We also perform statistical comparison of the $\Lambda$wDM scenario with the concordance model, where the terminology wDM corresponding to non-cold dark matter. \cref{tab:DIC} presents the best-fit $\Delta \chi^{2}_{\rm MAP}  = \chi^{2}_{\Lambda\rm wDM} - \chi^{2}_{\Lambda \rm CDM}$ along with their significance $N \sigma $, and $\Delta \rm DIC$. From the DESI alone, we obtain the 68\% CL constraints $w_{\rm dm} = -0.042^{+0.047}_{-0.024} $ and $\Omega_{\rm m} = 0.334^{+0.023}_{-0.040}$ by marginalizing over other cosmological parameters. The former excludes the canonical value of $w_{\rm dm} = 0$ at a statistical significance of $\approx 1.1\sigma$. The DESI data alone prefer the concordance model over the wDM scenario as we find $\Delta \rm DIC >0$ despite a better fit yielding $\Delta \chi^{2}_{\rm MAP} = -0.577$. Next, we jointly analyze DESI and SNIa data. SNIa data alone cannot determine the EoS for dark matter. Interestingly, the joint analysis of DESI+ SNIa tightly constrains the $w_{\rm dm}$ as suggested in \cref{tab:1sigma}. However the constraints on $w_{\rm dm}$ depend on the choice of SNIa dataset. In particular, the joint analysis of DESI and DESY5 yields a preference for non-vanishing $w_{\rm dm} = -0.084 \pm 0.035$ exceeding the 95\% CL. Additionally, we find a higher value of $\Omega_{\rm m} = 0.375^{+0.026}_{-0.032} $ compared to $\Lambda$CDM and $H_{0} = 75^{+10}_{-8}$. The large error bars on $H_0$ are due to the degeneracy between the Hubble parameter $H_0$ and the sound horizon at the drag epoch $z_{\rm d}$. The addition of CC data to DESI measurements breaks the $H_{0}\, r_{\rm d}$ degeneracy and does not affect the inference of a non-zero value of $w_{\rm dm}$ \footnote{The cosmic chronometers (CC) method relates the evolution of differential ages of passive galaxies at different redshifts without assuming any cosmological model \cite{Jimenez:2001gg}. We use the 32 measurements of Hubble parameter along with their covariance matrix}. The 2-dimensional posterior distributions of $h\,r_{\rm d} - w_{\rm dm}$ and $\Omega_{m}- w_{\rm dm}$ for DESI and DESI+ SNIa are shown in \cref{fig:wdm_posteriors} in the left and middle panels, respectively.
It is clear from \cref{fig:wdm_posteriors} that DESI and its combination with SNIa permit a wide range of posterior distribution of $\Omega_{m}$. The $\Lambda$wDM scenario is statistically preferred over $\Lambda$CDM, with $\Delta \chi^{2}_{\rm MAP} \in [-3.4, -7.9]$ corresponding to significance levels ranging from $1.85\sigma$ to $2.80\sigma$. We obtain a similar preference by evaluating the $\Delta$DIC as shown in \cref{tab:DIC}. The combination of $\rm DESI+DESY5$ shows a strong preference while $\rm DESI+PP$ and $\rm DESI+Union3$ indicate moderate preferences over the concordance model.

Next, we investigate how CMB measurements affect the inference of dark matter EoS $w_{\rm dm}$. The PL18 measurements alone find the $w_{\rm dm} = -0.0022 \pm 0.0020$, $H_{0} = 63.1\pm 3.9$ and $\Omega_{\rm m} = 0.383^{+0.051}_{-0.073}$. We observe that the geometrical degeneracy compensates for extreme values of $H_0$ and $\Omega_{m}$ through the negative contribution of dark matter EoS. Both PL18 and DESI exclude $w_{\rm dm}$ from $0$, but most of the posterior values of CMB inferred $w_{\rm dm}$ are not as widespread as those of DESI. The joint PL18+DESI analysis breaks the $H_0$-$\Omega_m$-$w_{\rm dm}$ geometrical degeneracy, yielding $w_{\rm dm} = 0.00077\pm0.00038$ ($\sim2\sigma$ from zero) and a higher $H_0$ than $\Lambda$CDM, though still in tension with $\rm{S}H_0\rm{ES}$ \cite{Riess:2016jrr,Riess:2020fzl}. This combination shows moderate preference over $\Lambda$CDM, with $\Delta\chi^2_{\rm MAP}=-5.60$ and $\Delta$DIC$=-3.32$. We also consider PL18 jointly with DESI and SNIa datasets, which even result in precise constraints on $w_{\rm dm}$. For instance, the most accurate constraint on $w_{\rm dm} = 0.00067 \pm 0.00037$ is derived from the PL18+DESI+Union3 combination. The $\Delta \chi^{2}_{\rm MAP}$ values are respectively $-3.82, -4.31$, and  $4.03$, indicating the preference of the $\Lambda$wDM scenario over $\Lambda$CDM at $\approx 2 \sigma$, for combining DESI with DESY5, PP and Union3, respectively. The similar preference can be corroborated from the last column of \cref{tab:DIC} using DIC except for the PP dataset. Hence, all combinations, including different SNIa measurements, consistently rule out $w_{\rm dm} = 0$ at least by $ 1.5 \sigma$. 

Finally, we conclude our analysis by considering the DESY3 ($3\times2$pt) in conjunction with DESI and SNIa data. First of all, unlike the SNIa measurements, $w_{\rm dm}$ can be inferred from DESY3 ($3\times2$pt) alone, as shown in \cref{tab:1sigma}. The right panel of \cref{fig:wdm_posteriors} shows the comparison of one dimensional posterior distribution of $w_{\rm dm}$ obtained from DESI alone and in combination with DESY5 and DESY3 likelihood. We observe that weak lensing measurements, even when combined with DESI and supernova data, yield a dark matter EoS consistent with $\Lambda$CDM, although with significantly weaker constraints compared to other datasets. Consistent with previous analyses, we find substantial evidence favoring our model over $\Lambda$CDM, as demonstrated in \cref{tab:DIC}. Assessing the compatibility between different dataset combinations is essential, as joint analyses yield significantly different constraints. To quantify the concordance between two datasets ($D_1$ and $D_2$), we compute the DIC-based statistic $\mathcal{I}$. \Cref{tab:I} presents the resulting $\log_{10} \mathcal{I}$ values for our dataset combinations. We find that combinations that favor $w_{\rm dm}<0$ exhibit stronger concordance, while those predicting $w_{\rm dm}>0$ show substantially lower compatibility.

The results obtained by combining DESI distance measurements with SNIa, CMB and DES-Y3 data clearly hint towards the preference of non-cold dark matter at varying level of significance. It is important to mention that such a preference is also present in DESI DR1. We have explicitly checked and found that our results are consistent with previous data release of DESI BAO. Furthermore, we also consider the SDSS BAO measurements from the final SDSS collaboration compilation comprising the eight effective redshift bins \cite{BOSS:2016wmc,BOSS:2016apd}. Both DESI and SDSS datasets focus on the same distance measurements and have similar constraining power. Therefore, it is necessary to assess our results obtained from the DESI measurements. We perform MCMC analysis for SDSS by adopting the same methodology as used in the case of DESI. We find no evidence for non-vanishing EoS for dark matter component from the SDSS alone or from the joint analysis of SDSS with any SNIa compilations used in this work. In particular, we obtain a weak bound on $w_{\rm dm} < -0.0275 $ using SDSS+DESY5. The analysis of other data such as Planck 2018 and DES-Y3 in conjunction with SDSS will not give much intuition in the inference of $w_{\rm dm} \neq 0$ as Planck and DES-Y3 alone can constrain $w_{\rm dm}$ precisely.

\begin{table}[!ht]
\centering

\setlength{\tabcolsep}{3pt}  
\renewcommand{\arraystretch}{1.0}  
\begin{tabular}{|p{4.5cm}|c|c|c|}
\hline
\hline
\textbf{Dataset} & $\bm{\Delta \chi^2_{\rm MAP}}$ & $\bm{\mathrm{N \sigma}}$ & $\bm{\Delta \mathrm{DIC}}$ \\
\hline
\hline
DESI            & $-0.5769$ & $0.75$  & $0.78$ \\
DESI+DESY5      & $-7.89$   & $2.80$  & $-5.87$ \\
DESI+PP         & $-3.43$   & $1.85$  & $-1.58$ \\
DESI+Union3     & $-4.08$   & $2.02$  & $-2.06$ \\
\hline
PL18            & $-0.10$    & $0.32$ & $-0.39$ \\
PL18+DESI       & $-5.60$   & $2.36$  & $-3.32$ \\
PL18+DESI+DESY5 & $-3.82$   & $1.95$  & $-2.32$ \\
PL18+DESI+PP    & $-4.31$   & $2.07$  & $-0.60$ \\
PL18+DESI+Union3& $-4.03$   & $2.0$  & $-2.36$ \\
\hline
DESI+DESY3 (3$\times$2pt) & $-3.06$   & $1.75$  & $-4.65$ \\
DESI+DESY3(3$\times$2pt)+DESY5 & $-2.69$   & $1.64$  & $-1.31$ \\
\hline
\hline
\end{tabular} 
\caption{Comparison of $\Lambda$wDM versus $\Lambda$CDM using different statistical tests: the maximum a posteriori $\chi^2$ difference ($\Delta \chi^2_{\rm MAP}$), significance level ($N\sigma$), and Deviance Information Criterion ($\Delta \mathrm{DIC}$). Negative $\Delta \chi^2_{\rm MAP}$ and $\Delta \mathrm{DIC}$ values favor the $\Lambda$wDM, with DESI+DESY5 showing the strongest preference ($\Delta \chi^2_{\rm MAP} = -7.89$, $2.8\sigma$ significance) while combinations with Planck (PL18) yield weaker evidence.}
\label{tab:DIC}
\end{table}

\begin{table}[!ht]
\centering

\setlength{\tabcolsep}{5pt}  
\renewcommand{\arraystretch}{1.0}  
\begin{tabular}{|p{5.0cm}|c|}
\hline
\hline
\textbf{Dataset ($D_1 \& D_2$)} & ${\log_{10} \,\,\mathcal{I}(\rm D_1 , D_2)}$  \\
\hline
\hline
DESI \& DESY5      & $0.94$ \\
DESI \& PP         & $1.41$ \\
DESI \& Union3     & $0.91$ \\
DESI \& DESY3(3$\times$2pt)     & $1.91$ \\
\hline
PL18 \& DESI       & $0.84$ \\
\hline
\end{tabular}
\caption{DIC-based statistic $\log \mathcal{I}(D_1, D_2)$ quantifying tension between cosmological datasets. Positive values indicate agreement between datasets, with higher values corresponding to stronger consistency. Combinations favoring $w_{\rm dm}<0$ show better concordance, while those predicting $w_{\rm dm}>0$ yield significantly lower compatibility.}
\label{tab:I}
\end{table}
We further investigate how the constituent tracers of the DR2 samples at different redshifts affect our findings of $w_{\rm dm} < 0$. To do so, we split the DESI samples into two subsets based on the effective redshift, separating at $z = 1.1$. Interestingly, we see that the DESI DR 2 samples with $z>1.1$ still provide $w_{\rm dm} <0$ but consistent with $w_{\rm dm} = 0$ at varying CL for DESI and SNIa measurements, as shown in \cref{tab:binned}. The samples containing distances up to $z<1.1$ are not able to constrain dark matter's EoS. Although, upon a close inspection it turns out that 90\% of the posterior values of $w_{\rm dm}$ disfavor the possibility of cold dark matter.

\begin{table}[!ht]
\centering
\setlength{\tabcolsep}{4pt}
\renewcommand{\arraystretch}{1.2}
\begin{tabular}{|l|c|c|c|}
\hline
\hline
Dataset & $\bm{w_{\mathrm{dm}}}$ & $\bm{\Omega_{m}}$ & $\bm{h\,r_{\mathrm{d}}}$ \\
\hline
\hline
DESI($z>1.1$) & $-0.049^{+0.070}_{-0.022}$ & $0.358^{+0.027}_{-0.089}$ & $99.2^{+3.8}_{-2.7}$ \\
DESI+DESY5 & $-0.064^{+0.042}_{-0.033}$ & $0.376^{+0.026}_{-0.031}$ & $97.8 \pm 1.4$ \\
DESI+PP & $-0.043^{+0.039}_{-0.024}$ & $0.347^{+0.023}_{-0.030}$ & $99.2 \pm 1.5$  \\
DESI+Union3 & $-0.068^{+0.055}_{-0.035}$ & $0.380^{+0.033}_{-0.053}$ & $97.8 \pm 1.8$  \\
\hline
DESI($z<1.1$) & $< -0.151$ & $0.454 \pm 0.090$ & $99.1 \pm 1.5$ \\
DESI+DESY5 & $< -0.220$ & $0.487^{+0.062}_{-0.055}$ & $98.8 \pm 0.8$ \\
DESI+PP & $-0.184^{+0.065}_{-0.120}$ & $0.425^{+0.052}_{-0.067}$ & $99.7 \pm 0.8$  \\
DESI+Union3 & $< -0.217$ & $0.486^{+0.083}_{-0.056}$ & $98.6 \pm 1.1$  \\
\hline
\hline
\end{tabular} 
\caption{68\% constraints on $w_{\mathrm{dm}}$, $\Omega_m$, and $h\,r_{\mathrm{d}}$ from DESI BAO binned at $z>1.1$ (consistent with $\Lambda$CDM) and $z<1.1$ (favors $w_{\mathrm{dm}}<0$), combined with DES-Y5, PP, and Union3 supernovae.}
\label{tab:binned}
\end{table}

\begin{figure}[!ht] 
    \includegraphics[scale=0.55]{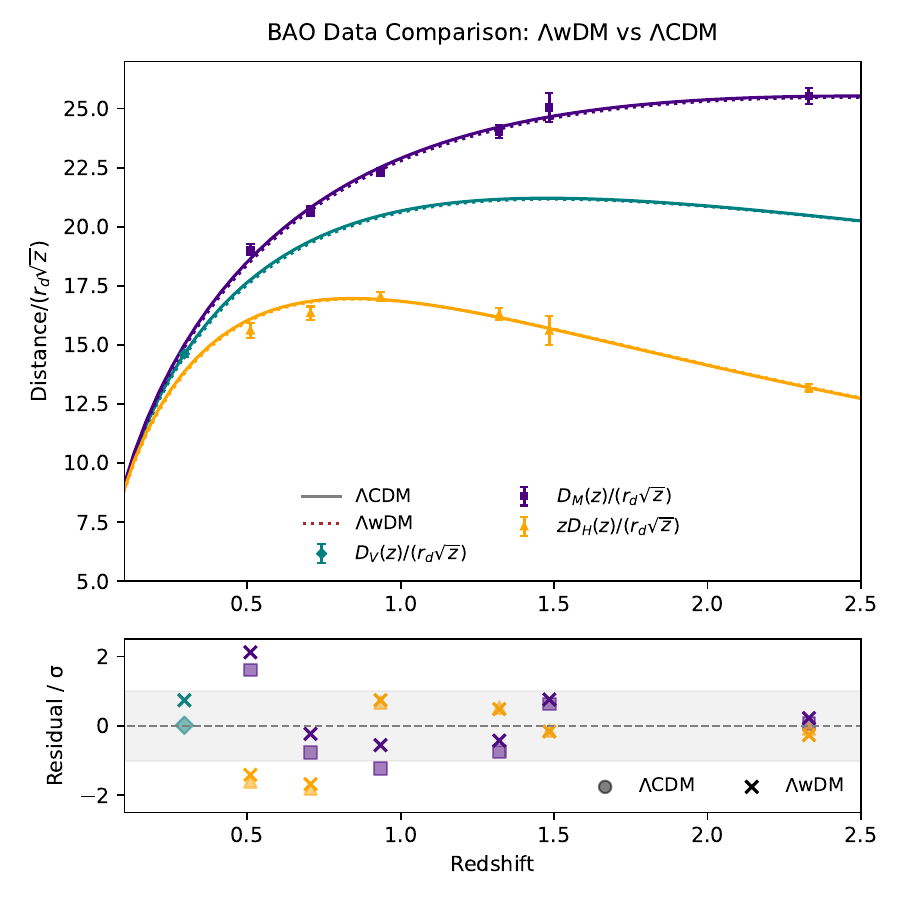}
    \caption{\textit{Upper Panel:} Comparison of rescaled distance-redshifts relations for $\Lambda$CDM (solid) and $\Lambda$wDM (dashed) model using the best-fit values inferred from DESI+Planck 2018 joint analysis. We also include the observational constraints with $\pm 1 \sigma$ uncertainty. Different type of distance measurements are indicated by different colors given in the legend. \textit{Lower Panel:} The deviation of theoretical rescaled distances from the corresponding data-points in terms of $\sigma$. The predictions for $\Lambda$CDM and $\Lambda$wDM are shown by filled and cross shaped markers, respectively. The shaded region highlights the $\pm1\sigma$ range for reference.}
    \label{fig:distances}
\end{figure}

To get a better understanding of the role played by DESI data, we also compare the theoretical distance predictions for $\Lambda$wDM and $\Lambda$CDM model with the observed distances probed by BAO measurements as shown in \cref{fig:distances}. The lower panel of \cref{fig:distances} demonstrates that the theoretical distance deviations obtained using the $\Lambda$wDM model exhibit significant discrepancies with the concordance model for redshifts $z < 1.1$. In contrast, for $z > 1.1$, the results show strong agreement with the concordance model. Similar conclusions are also drawn from the analysis of the Binned DESI data discussed previously. Moreover, we observe that the primary contributor to the discrepancy between the models at $z < 1.1$ is the distance measure $D_{\rm M}(z) / (r_{\rm d} \sqrt{z})$.

Finally, we discuss the implications of the preference for $ w_{\rm dm} < 0 $ beyond the context explored in this work. As argued by the DESI collaboration, allowing for time-varying dark energy parametrized by the Chevallier-Polarski-Linder (CPL) form \cite{Chevallier:2000qy, Linder:2002et} yields a strong statistical preference for such a scenario. A non-zero equation of state (EoS) for dark matter could naturally arise from interactions between the dark matter (DM) and dark energy (DE) sectors \cite{He:2008tn,Gavela:2009cy,Clemson:2011an}. Indeed, multiple studies have reported evidence for a non-zero DM-DE coupling parameter \cite{Yang:2025ume,Pan:2022qrr,Hou:2022rvk,Pan:2025qwy,Giare:2024smz,Silva:2025hxw,Kessler:2025kju,}. Alternatively, One can also envisage a scenario where DM and DE are described by a single fluid component \cite{Anagnostopoulos:2019myt,Perkovic:2019vxm,Koutsoumbas:2017fxp}. Such Unified DM-DE (UDM) models have been shown to provide promising avenues for resolving both the Hubble tension and large-scale structure anomalies \cite{Frion:2023xwq,PhysRevD.111.063522}.

In this letter, we have derived constraints on the dark matter equation of state parameter $w_{\rm dm}$ using the latest DESI BAO measurements, both independently and in combination with Planck, DES, and supernovae data. Our analysis reveals a robust preference for non-cold dark matter across multiple dataset combinations involving DESI. Notably, we observe a tension between background-based measurements (including SNIa compilations, cosmic chronometers, and BAO) and observations involving perturbations (such as Planck and DES Y3). The combination DESI+DESY5 yields $w_{\rm dm} = -0.084\pm0.035$, excluding the cold dark matter scenario ($w_{\rm dm}=0$) at $2.4 \sigma$ significance, while PL18+DESI gives $w_{\rm dm} = 0.00077\pm0.00038$, differing from the canonical value at $2\sigma$. Although these estimates differ numerically, they both consistently suggest the existence of non-cold dark matter. We highlight that the preference for negative $w_{\rm dm}$ in DESI is primarily driven by low-redshift BAO measurements ($z<1.1$), whereas higher redshift BAO data ($z>1.1$) remain consistent with $\Lambda$CDM. Intriguingly, this evidence for $w_{\rm dm}<0$ is more pronounced in DESI, as previous surveys like SDSS were only able to place weak constraints such a preference. For all dataset combinations considered, we observe significant improvements over the standard $\Lambda$CDM paradigm.

Our results demonstrate that the DESI dataset opens new avenues for investigating viable dark matter models via late-time physics. While this study focuses on the implications of DESI measurements for dark matter properties inferred from geometric distances, it is important to confirm these findings through full-shape modeling of the power spectrum, incorporating the effects of redshift-space distortions.
  
\textit{\textbf{Note Added:-}} During the finalization of this work, \cite{Yang:2025ume,Pan:2025qwy} appeared on arXiv, investigating dark matter properties with DESI data while incorporating interactions in the dark matter–dark energy (DM-DE) sector. In contrast, our analysis focuses on minimal extensions of the standard cosmological model and combines complete CMB power spectra measurements with DES-Y3 $3\times2$ pt data. Unlike their findings, we observe a preference for the extended cosmological scenario over the $\Lambda$CDM.

\textit{\textbf{Acknowledgments}}-- We thank Prof. Ido Ben-Dayan and Prof. Sukanta Panda for their valuable comments. We also acknowledge the Open U HPC Center at Open University of Israel for providing computing resources that have contributed to the research results reported in this paper. We acknowledge Prof. Sunny Vagnozzi for providing the Python module utilized to create \cref{fig:distances}.

	\newpage

	\newpage
	\pagebreak
	\appendix
	\onecolumngrid

\newpage	
\bibliography{reference.bib}	
\newpage
\section*{Comparison of DESI and SDSS cosntraints}
In this extended analysis, we build upon the study presented in the main text by substituting the DESI dataset with the Baryon Acoustic Oscillation (BAO) measurements obtained from the final compilation of the SDSS collaboration. This dataset encompasses a comprehensive set of observations across eight distinct redshift intervals, providing an alternative yet consistent framework for examining the same cosmological parameters. The specific redshift ranges and corresponding measurements are detailed in \cite{eBOSS:2020yzd}. As in the previous analysis, we incorporate the BAO dataset in combination with additional cosmological probes, including the DESY5 galaxy clustering data, PantheonPlus (PP) measurements, and the Union 3 compilation of Type Ia supernova observations, to ensure a comprehensive and robust constraint on the model parameters.

We summarize the observational constraints derived from our analysis in \cref{tab:sdss}. Notably, we do not obtain any robust constraints on the dark matter equation-of-state parameter, $w_{dm}$, regardless of the combination of datasets considered. However, our results consistently indicate a higher value of the matter density parameter, $\Omega_m$, and a lower value of the product $h\,r_d$, compared to our analysis with the DESI.

\begin{table}[!ht]
\centering
\setlength{\tabcolsep}{10pt}
\renewcommand{\arraystretch}{1.2}
\begin{tabular}{|l|c|c|c|c|}
\hline
\hline
Dataset & $\bm{H_0}$ & $\bm{w_{dm}}$ & $\bm{\Omega_{m}}$ & $\bm{h\,r_{\mathrm{d}}}$ \\
\hline
\hline
SDSS & $75^{+20}_{-10}$ & $< -0.0256 $ & $0.413^{+0.058}_{-0.076}$ & $97.5 \pm 2.2$ \\
SDSS+DESY5 & $76^{+20}_{-10}$ & $<-0.0321$ & $0.384^{+0.026}_{-0.030}$ & $98.5^{1.00}_{0.92}$ \\
SDSS+PP & $77^{+20}_{-9}$ & $<-0.0290$ & $0.363^{+0.026}_{-0.032}$ & $99.2 \pm 1.0$  \\
SDSS+Union3 & $74^{+20}_{-10}$ & $< - 0.0201$ & $0.390^{+0.032}_{-0.040}$ & $98.1^{1.10}_{0.96}$  \\
\hline
\hline
\end{tabular} 
\caption{68\% constraints on $w_{\mathrm{dm}}$, $\Omega_m$, and $h\,r_{\mathrm{d}}$ from SDSS BAO alone and combined with DES-Y5, PP, and Union3 supernovae.}
\label{tab:sdss}
\end{table}

\end{document}